\documentclass[12pt]{article}
\newcommand{\beq}{\begin{equation}}
\newcommand{\eeq}{\end{equation}}
\newcommand{\beqa}{\begin{eqnarray}}
\newcommand{\eeqa}{\end{eqnarray}}
\newcommand{\beqar}{\begin{eqnarray*}}
\newcommand{\eeqar}{\end{eqnarray*}}

\newcommand{\eps}{\epsilon}

\newcommand{\inn}{\!\cdot\!}

\newcommand{\z}{\zeta}

\newcommand{\eg}{{\it e.g.,}\ }
\newcommand{\ie}{{\it i.e.,}\ }
\newcommand{\labell}[1]{\label{#1}} 
\newcommand{\reef}[1]{(\ref{#1})}
\newcommand\prt{\partial}

\newcommand\cF{{\cal F}}

\newcommand\cA{{\cal A}}

\newcommand\cM{{\cal M}}

\newcommand\cB{{\cal B}}

\newcommand\tq{{\tilde q}}
\newcommand\tF{{\widetilde F}}

\newcommand\Tr{{\rm Tr}}

\begin{document}

\vspace*{1cm}

\begin{center}
{\bf \Large S-duality of tree-level S-matrix elements \\ \vspace*{0.5cm}  in D$_3$-brane effective action}

\vspace*{1cm}

{H. Babaei-Aghbolagh\footnote{hossein.babaei66@gmail.com} } and Mohammad R. Garousi\footnote{garousi@um.ac.ir}  \\
\vspace*{1cm}
{ Department of Physics, Ferdowsi University of Mashhad,\\ P.O. Box 1436, Mashhad, Iran}
\\
\vspace{2cm}

\end{center}

\begin{abstract}
\baselineskip=18pt

Recently it has been speculated  that the S-matrix elements on the world volume of D$_3$-branes  should satisfy the ward identity corresponding to the S-duality transformations. For a single brane and in the low energy limit, this indicates that in the  abelian Dirac-Born-Indeld and Chern-Simons actions the combination of  contact terms and   tree-level massless poles in an $n$-point function should satisfy the S-duality. 
In this paper we examine in details the S-matrix element of three gauge bosons and one two-form and the S-matrix element of six gauge bosons   in favor of the above proposal.
 \end{abstract}

\vfill
\setcounter{page}{0}
\setcounter{footnote}{0}
\newpage


\section{Introduction } \label{intro}
It is known that the  type IIB superstring theory  is invariant  under S-duality \cite{Font:1990gx,Rey:1989xj,Sen:1994fa,Sen:1994yi,Schwarz:1993cr,Hull:1994ys}. This symmetry  should be carried by  the S-matrix elements through the associated Ward identity \cite{Garousi:2011we}.  That is, the S-matrix elements should be invariant under linear S-duality transformations on the external states and  nonlinear S-duality transformation on the background fields. In particular, the S-matrix elements on the world volume  of D$_3$-branes should be invariant under nonlinear S-duality transformations on the coupling constant and on the background R-R scalar field. This requires, in general, to combine the tree-level amplitudes with loop  and  non-perturbative amplitudes \cite{Green:1997tv} - \cite{Garousi:2011fc}.

The   world volume theory of a single D$_3$-brane at low energy is given by the Dirac-Born-Infeld-Chern-Simons (DBICS) action \cite{Leigh:1989jq,Bachas:1995kx,Polchinski:1995mt,Douglas:1995bn}.  This action is not invariant under the S-duality, however, its equations of motion and its energy-momentum tensor are invariant under the S-duality \cite{Gibbons:1995ap}. There are proposals for the $SL(2,R)$-covariant form of the D$_3$-brane action \cite{Cederwall:1997ab,Bergshoeff:2006gs}. In this paper, we would like to study the S-dual Ward identity of the S-matrix elements in the DBICS theory. This theory is not renormalizable, hence, the degree of momentum  of an S-matrix element   at tree-level is different from the degree of momentum  at one  loop-level,  the degree of momentum at one loop-level is different from the degree of momentum at two loop-level, and so on. On the other hand, the standard S-duality transformations do not receive $\alpha'$ correction. As a result, the S-dual Ward identity in the DBICS theory can not relate the tree-level amplitudes to   loop-level amplitudes.  Instead, the  Ward  identity should   leave invariant the tree-level amplitudes. It should relate the one loop-level amplitudes in this theory to the appropriate higher derivative corrections of the   DBICS theory and to the appropriate non-perturbative  amplitudes. Similarly for higher loop-level amplitudes.
In this paper, we would like to show by explicit calculations    that the combination of tree-level massless poles and contact terms in the DBICS theory satisfies the S-dual Ward identity. 

The outline of the paper is as follows: We begin in section 2 by reviewing the low energy effective action of a single D$_3$-brane. In section 3, we calculate the scattering amplitude of three gauge bosons and one two-form in the presence of the  background  dilaton and R-R scalar fields. In subsection 3.1, we show that the amplitude satisfies the Ward identity corresponding to $SL(2,R)$ transformations. In section 4, we calculate the S-matrix element of six gauge bosons and show that the amplitude satisfies the S-dual Ward identity.

 \section{D-brane effective action}

The bosonic massless fields on the world volume  of D-branes are the world volume gauge boson and the transverse scalar fields. When the derivatives of the world volume  field strengths and the derivative of bulk fields are small compare to  $\alpha'$, the dynamics of the D-branes of type II superstring theories is well-approximated by the effective world-volume field theory  which consists of the  Dirac-Born-Infeld (DBI) action and the Chern-Simons (CS) part. 
The DBI action  describes the dynamics of the brane in the presence of  NS-NS background fields, which  can be found by requiring its  consistency with the  nonlinear T-duality \cite{Leigh:1989jq,Bachas:1995kx}. On the other hand,  the CS part describes the coupling of D-branes to the R-R potentials \cite{Polchinski:1995mt,Douglas:1995bn}. The DBICS action in the Einstein  frame for 
 single  D$_3$-brane  is\footnote{ Our index convention is that the Greek letters  $(\mu,\nu,\cdots)$ are  the indices of the space-time coordinates, the Latin letters $(a,d,c,\cdots)$ are the world-volume indices and the letters $(i,j,k,\cdots)$ are the normal bundle indices.} 
\beqa
S_{D3}&=&-T_{D3}\int d^4x\sqrt{-\det(g_{ab}+e^{-\phi/2}B_{ab})}\labell{DBICS}\\
&&+T_{D3}\int [C^{(4)}+C^{(2)}\wedge B+\frac{1}{2}C^{(0)}\wedge B\wedge B ]\nonumber
\eeqa
All the bulk fields in the action are pull-back onto  the world-volume of D-brane, \eg
\beqa
g_{ab}&=&g_{\mu\nu}\frac{\prt X^{\mu}}{\prt \sigma^a}\frac{\prt X^{\nu}}{\prt \sigma^b}
\eeqa
where $X^a=\sigma^a$  in the static gauge,  and the transverse scalar fields are all other components of $X^{\mu}$, \ie $X^i=2\pi\alpha'\Phi^i$. The abelian gauge field can be added to the action as $B\rightarrow\cF$ where $\cF= B+2\pi\alpha'F$. This makes the action to be invariant under the B-field gauge transformation.  This action is also  invariant under the following R-R gauge transformations:
\beqa
\delta C=d\Lambda+H\wedge \Lambda \labell{RR}
\eeqa
  where  $C=C^{(4)}+C^{(2)} $, $\Lambda= \Lambda^{(3)}+\Lambda^{(1)}$ and $H$ is the field strength of $B$, \ie $H=dB$.

   It has been shown in \cite{Gibbons:1995ap} that the equations of motion and energy-momentum tensor of the above action are invariant under the $SL(2,R)$ transformation. However, the action   itself  is not invariant under  this   transformation. This is resulted from the fact that the gauge field in this action is the massless mode of the fundamental open string propagating on the world-volume of D$_3$-brane. Under the S-duality, the fundamental string transforms to the D-string whose massless mode is given by another gauge field, \ie $\tF$, which is the dual of the Born-Infeld gauge field $F$. There are proposals for the $SL(2,R)$-covariant form of the D$_3$-brane action which include both Born-Infeld gauge field and its dual \cite{Cederwall:1997ab,Bergshoeff:2006gs}. A complex $SU(1,1)$-base $q_{\alpha}$   has been considered in \cite{Bergshoeff:2006gs}. By rotating  this doublet at the same time that rotating the world volume and bulk fields,  an    $SU(1,1)$-covariant family of actions  has been found in  \cite{Bergshoeff:2006gs}, \ie
\beqa
S_{D3}&=&-T_{D3}\int d^4x\sqrt{-\det\left(g_{ab}+\frac{q_{\alpha}\cF^{\alpha}_{ab}}{(q_{\alpha}\cM^{\alpha\beta}q_{\beta})^{1/2}}\right)}+\cdots\labell{pq} 
\eeqa
where dots represent the Chern-Simons part which includes another doublet $\tq_{\alpha}$, and the matrix $\cM$, in the real $SL(2,R)$-base $q_{\alpha'}$, appears in \reef{M}. In this base,  the DBI action \reef{DBICS} corresponds to  $q_{\alpha'}=(0, -1)$ and $\tq_{\alpha'}=(1,0)$. Under the S-duality, $q_{\alpha'}\rightarrow \tq_{\alpha'}$ \cite{Bergshoeff:2006gs}. The DBI action then transforms under the S-duality to
 \beqa
S_{D3}&=&-T_{D3}\int d^4x\sqrt{-\det\left(g_{ab}+ \frac{C_{ab}+2\pi\alpha' \tF_{ab}}{(e^{-\phi}+C_{(0)}^2e^{\phi})^{1/2}}\right)}+\cdots \labell{DBIS} 
\eeqa  
Using the actions \reef{DBICS} and \reef{DBIS}, one may calculate various tree-level S-matrix elements. Their contact terms which are given by the appropriate terms  in \reef{DBICS} and \reef{DBIS}, are related to each other by the S-duality, and  similarly the massless poles in  \reef{DBICS}   are related to the corresponding massless poles in the above theory by the S-duality. In this way one can calculate an $SL(2,R)$-covariant family of   S-matrix elements.
   
In this paper, however, we are interested in studying to what extend  the S-matrix elements in one of the above $SL(2,R)$-covariant family of actions are  $invariant$ under the $SL(2,R)$ transformations.  In particular, we are interested in studying the symmetry of the S-matrix elements in \reef{DBICS} which includes only one Born-Infeld field $F$, under the S-duality transformation. Under the S-duality, the gauge field strength $F$ in \reef{DBICS} transforms to the nonlinear combination of $F$ \cite{Gibbons:1995ap}, so it is nonsense to study the invariant of an S-matrix element with fix number of  asymptotic   gauge bosons. However, if the theory is invariant under the S-duality, one expects their S-matrix elements to satisfy the Ward identity corresponding to the S-duality which is a linear transformation on the asymptotic   states \cite{Garousi:2011we}.

\section{Three gauge bosons and one two-form amplitudes}
 
Using the action \reef{DBICS}, one can calculate the scattering amplitude of three gauge bosons and one two-form in the presence of the background dilaton and R-R scalar  fields. The two-form can be either the antisymmetric B-field or the R-R two-form. When the two-form is the R-R two-form, the amplitude has only massless pole which is given by the following Feynman rule:
\beqa
\cA(C_4^{(2)})&=&V^a(F_1,F_2,F_3, A)G_{ab}(A)V^b(A,C_4^{(2)})\labell{amp1}
\eeqa
where $F_1,\, F_2,\, F_3$ are the polarizations of the external gauge bosons, $C_4^{(2)}$ is the polarization of the external R-R two-form and $A$ is the off-shell  gauge field propagating between the two vertices. The vertices and the gauge boson propagator can be read from \reef{DBICS} to be 
\beqa
V^a(F_1,F_2,F_3, A)&\sim& -e^{-2\phi_0}\bigg[k\inn F_1\inn F_2\inn F_3^a-\frac{1}{4} k\inn F_1^a\Tr(F_2\inn F_3)\bigg]+P(1,2,3) \labell{ver}\nonumber\\
 V^b(A,C_4^{(2)})&\sim&k_4\inn(*C_4^{(2)})^b  \nonumber\\
G_{ab}(A)&\sim&-\frac{e^{\phi_0}\eta_{ab}}{k\inn k}\labell{poro}
\eeqa
where $(*C_4^{(2)})_{ab}=\frac{1}{2}\eps_{abcd}(C_4^{(2)})^{cd}$. In above equations, $P(1,2,3)$ stands for  the other five permutations of $1,2,3$, and $k$ is the off-shell momentum in the propagator which is $k=-k_4=k_1+k_2+k_3$. We have ignored the factors of $(2\pi\alpha')$ and $T_{D_3}$. 

Replacing \reef{poro} in the amplitude \reef{amp1}, one finds the following result:
\beqa
\cA ( C_4^{(2)} )\sim-\frac{e^{- \phi _0}}{k_4\inn k_4}\bigg[ k_4\inn F_1\inn F_2\inn F_3\inn *C_4^{(2)}\inn k_4 -\frac{1}{4} k_4\inn F_1\inn *C_4^{(2)}\inn k_4 \Tr(F_2\inn F_3)\bigg]  +P(1,2,3) \labell{CFFF}
\eeqa
The amplitude satisfies the Ward identity corresponding to the abelian gauge symmetry and to the R-R gauge symmetry, \eg if one replaces $C_4^{(2)}\rightarrow k_4\wedge\z_4^{(1)}$ where $\z_4^{(1)}$ is an arbitrary one-form, the amplitude becomes zero.

When the two-form is the B-field, the amplitude has   massless pole as well as contact term which are given by the following Feynman rule:
\beqa
\cA(B_4)&=&V^a(F_1,F_2,F_3, A)G_{ab}(A)V^b(A,B_4)+V(F_1,F_2,F_3, B_4)\labell{amp2}
\eeqa
where $B_4$ is the polarization of the external B-field. The vertex $V^b(A,B_4)$ and the contact term $V(F_1,F_2,F_3, B_4)$ can be read from \reef{DBICS} to be
\beqa
V^b(A,B_4)&\sim& -e^{-\phi_0}k_4\inn B_4^b+C_0\,\, k_4\inn *B_4^b \labell{ver2}\\
V(F_1,F_2,F_3, B_4)&\sim&\frac{1}{2}e^{-2 \phi _0} \left( \Tr(B_4\inn F_1 \inn F_2 \inn F_3)-\frac{1}{4}  \Tr(B_4\inn F_1)\Tr(F_2\inn F_3)\right)+P(1,2,3)\nonumber
 \eeqa
Note that the dot means the contraction of the world volume indices, so even though $(k_4)_{\mu}B_4^{\mu\nu}=0$, the expresion $k_4\inn B_4^b$ in the first line is not zero. Replacing the vertices and propagator in \reef{amp2}, one finds the following result: 
\beqa
\cA ( B_4 ) &\sim&-\frac{1}{2}e^{-2 \phi _0} \bigg[  \Tr(B_4\inn F_1 \inn F_2 \inn F_3)-\frac{1}{4}  \Tr(B_4\inn F_1)\Tr(F_2 \inn F_3)\bigg]\nonumber\\
&&\qquad+\frac{e^{-2 \phi _0} }{k_4\inn k_4}\bigg[  k_4\inn F_1\inn F_2\inn  F_3\inn B_4\inn k _4-\frac{1}{4}  k_4\inn F_1\inn B_4\inn k _4 \Tr(F_2 \inn F_3)  \bigg]\labell{BFFF}\\
&&+\frac{C_0e^{-\phi_0 }}{k_4\inn k_4} \bigg[k_4\inn F_1\inn F_2\inn F_3\inn *B_4\inn k_4 -\frac{1}{4}k_4\inn F_1\inn *B_4\inn k_4\Tr(F_2\inn F_3)\bigg]+ P(1,2,3)\nonumber
\eeqa
One can easily check that the amplitude satisfies the Ward identity corresponding to the  B-field gauge symmetry, \ie if one replaces $B_4^{ab}\rightarrow k_4^a\z_4^b-k_4^b\z_4^a$, the amplitude vanishes.

\subsection{S-dual Ward identity}

To study the amplitudes \reef{CFFF} and \reef{BFFF} under the Ward identity corresponding to the global S-duality, one has to find the linear S-duality transformations on the external states and nonlinear S-duality transformations on the background fields \cite{Garousi:2011we,Garousi:2011vs}. The S-duality transformation on the  two-forms is a linear transformation,  and on the gauge field is a nonlinear transformation\cite{Gibbons:1995ap,Tseytlin:1996it,Green:1996qg}. The S-duality transformation on the two-forms and the linearized S-duality transformation on the gauge field are the following \cite{Gibbons:1995ap,Tseytlin:1996it,Green:1996qg}:
\beqa
\mathcal{B}\longrightarrow (\Lambda^{-1})^T\mathcal{B}\,,\,\mathcal{F}\longrightarrow (\Lambda^{-1})^T\mathcal{F} &;&\Lambda\in\, {SL(2,R)}\labell{BF}
\eeqa
where the doublets $\mathcal{B}$ and $\mathcal{F}$ are
\beqa
\cB\equiv\left(
\begin{array}{c}
B \cr  \
C^{(2)}
\end{array}
\right)\,\,\,,\,\,\,
\cF\equiv\left(
\begin{array}{c}
*F \cr  \
 e^{-\phi_0}F-C_0*F
\end{array}
\right)
\eeqa
The S-duality transformation on the background fields $\phi_0$ and $C_0$ is  \cite{Gibbons:1995ap}
\beqa
{\cal M}_0\rightarrow \Lambda {\cal M}_0\Lambda ^T\labell{MM}
\eeqa
where the matrix $\cM_0$ is 
\beqa
 {\cal M}_0=e^{\phi_0}\pmatrix{|\tau_0|^2&C_0 \cr 
C_0&1}\labell{M}
\eeqa
and $\tau_0=C_0+ie^{-\phi_0}$. Since the dilaton and the R-R scalar appear only as the background fields in the amplitudes  \reef{CFFF} and \reef{BFFF}, we keep the nonlinear form of \reef{MM} in studying the S-dual Ward identity.

To demonstrate that the amplitudes  \reef{CFFF} and \reef{BFFF} satisfy the S-dual Ward identity, we have to show that they are invariant under the above transformations. To this end, we consider the following terms:
\beqa
 (*\cF^T)_a{}^c\cM_0\cB_{cb}&=&e^{-\phi_0}F_a{}^cB_{cb}-(*F)_a{}^cC_{cb}^{(2)}-C_0(*F)_a{}^cB_{cb} \nonumber\\
(\cF_1^T)_a{}^c\cM_0\cF_{2cb}&=& e^{-\phi_0}[(*F_1)_a{}^c(*F_2)_{cb}+F_{1a}{}^cF_{2cb}] \labell{invar}
\eeqa
Using the transformations \reef{BF} and \reef{MM}, one can easily verify that the left-hand sides are invariant under the S-duality. Using these S-duality invariant terms and using    the following identity:
\beqa
\eps^{abcd}\eps^{efgh}=-\left|\matrix{\eta^{ae}& \eta^{af}& \eta^{ag}&\eta^{ah}\cr 
\eta^{be}& \eta^{bf}& \eta^{bg}&\eta^{bh}\cr
\eta^{ce}& \eta^{cf}& \eta^{cg}&\eta^{ch}\cr
\eta^{de}& \eta^{df}& \eta^{dg}&\eta^{dh}}\right|
 \labell{zero2n} 
 \eeqa 
 one can construct the following S-duality invariant contact term:
\beqa
&&\Tr(\cF_1^T\cM_0\cF_2*\cF_3^T\cM_0\cB_4)=\labell{S1}\\
&& -e^{-2 \phi _0}\bigg[ \Tr(B_4\inn F_1\inn F_2 \inn F_3)+\Tr(B_4\inn F_2\inn F_1\inn F_3)-\frac{1}{2}  \Tr(B_4\inn F_3)\Tr(F_1\inn F_2)\bigg]\nonumber\\
&&-C_0e^{- \phi _0} \bigg[ \Tr(*B_4\inn F_1 \inn F_2 \inn F_3)+\Tr(*B_4\inn  F_2 \inn F_1\inn  F_3)-\frac{1}{2}  \Tr(*B_4\inn F_3)\Tr(F_1 \inn F_2)\bigg]\nonumber\\
&& -e^{- \phi _0} \bigg[ \Tr(*C_4^{(2)} \inn F_1 \inn F_2 \inn F_3)+\Tr(*C_4^{(2)}\inn  F_2 \inn F_1\inn F_3)-\frac{1}{2}  \Tr(*C_4^{(2)}\inn F_3)\Tr(F_1 \inn F_2)\bigg]\nonumber
\eeqa
where the traces are over the world volume indices, and the following S-duality invariant massless pole:
 \beqa
&&\frac{k_4\inn\cF_1^T\cM_0\cF_2 *\cF_3^T\cM_0\cB_4\inn k_4}{k_4\inn k_4}=\labell{S2} \\ 
&&\frac{e^{-2\phi_0 }}{k_4\inn k_4}\bigg[ - k_4\inn B_4\inn F_3\inn F_1\inn F_2\inn k_4- k_4\inn B_4\inn F_3\inn F_2\inn F_1\inn k_4+\frac{1}{2}  k_4\inn B_4\inn F_3\inn k_4 \Tr(F_1\inn F_2)\bigg]+\nonumber\\
&&\frac{e^{-\phi_0 }}{k_4\inn k_4} \bigg[  k_4\inn *C_4^{(2)}\inn F_3\inn F_2\inn F_1\inn k_4+ k_4\inn *C_4^{(2)}\inn F_1\inn F_2\inn F_ 3\inn k_4-\frac{1}{2}  k_4\inn *C_4^{(2)}\inn F_1\inn k_4 \Tr(F_2\inn F_3)\nonumber\\
&&\qquad\quad - k_4\inn k_4\Tr(F_1\inn F_2\inn F_3 \inn *C_4^{(2)})+\frac{1}{4}  k_4\inn k_4 \Tr(F_1\inn *C_4^{(2)}) \Tr(F_2\inn F_3)\nonumber\\
&&\qquad\quad -\frac{1}{2}  k_4\inn F_2\inn F_ 3\inn k_4 \Tr(F_1\inn *C_4^{(2)})+\frac{1}{2}  k_4\inn F_1\inn F_2\inn k_4\Tr(F_3\inn *C_4^{(2)})\nonumber\\
&&\qquad\quad - k_4\inn F_1\inn F_2\inn *C_4^{(2)}\inn F_ 3\inn k_4+ k_4\inn F_2\inn F_3\inn *C_4^{(2)}\inn F_1\inn k_4\bigg]+\nonumber\\
&&\frac{C_0e^{-\phi_0 }}{k_4\inn k_4} \bigg[ k_4\inn *B_4\inn F_3\inn F_2\inn F_1\inn k_4+ k_4\inn *B_4\inn F_1\inn F_2\inn F_ 3\inn k_4-\frac{1}{2}  k_4\inn *B_4\inn F_1\inn k_4 \Tr(F_2 \inn F_3)\nonumber\\
&& \qquad\quad- k_4\inn k_4 \Tr(F_1\inn F_2\inn F_3\inn *B_4)+\frac{1}{4}  k_4\inn k_4 \Tr(F_1\inn *B_4) \Tr(F_2 \inn F_3)\nonumber\\
&& \qquad\quad-\frac{1}{2}  k_4\inn F_2\inn F_ 3\inn k_4 \Tr(F_1\inn *B_4)+\frac{1}{2}  k_4\inn F_1\inn F_2\inn k_4\Tr(F_3 \inn *B_4)\nonumber\\
&& \qquad\quad- k_4\inn F_1\inn F_2\inn *B_4\inn F_ 3\inn k_4+ k_4\inn F_2\inn F_3\inn *B_4\inn F_1\inn k_4\bigg]\nonumber
\eeqa
Note that the right-hand side of above equation has both massless poles and contact terms, \ie the terms which have coefficient $k_4\inn k_4$ are contact terms. The contact terms have the same structure as the contact terms in the second and the third lines of \reef{S1}. On the other hand, the scattering amplitudes \reef{CFFF} and \reef{BFFF} do not have the contact terms that appear in the second and the third lines of \reef{S1}. Hence, the two S-duality invariants \reef{S1} and \reef{S2} should be combined with appropriate coefficients to cancel these undesirable contact terms, and produce the contact terms and the massless poles in the scattering amplitudes \reef{CFFF} and \reef{BFFF}. In fact, one can write the combination of these amplitudes, \ie $A=\cA(C_4^{(2)})+\cA(B_4)$ as 
\beqa
A\sim\frac{1}{4}\Tr(\cF_1^T\cM_0\cF_2*\cF_3^T\cM_0\cB_4)-\frac{1}{2}\frac{k_4\inn\cF_1^T\cM_0\cF_2 *\cF_3^T\cM_0\cB_4\inn k_4}{k_4\inn k_4}+ P(1,2,3)  \labell{A1}
\eeqa
Hence, the tree-level S-matrix element of three gauge bosons and one two-form satisfies the Ward identity corresponding to the gauge symmetries and to the S-duality. Note that there is no need to add loop or nonperturbative effects to satisfy the S-dual Ward identity. The   loop amplitudes in the DBICS theory are in higher order of momentum, so the S-dual Ward identity does not allow to  add them  to the tree-level amplitudes.  
 
We have also calculated the scattering amplitude of one gauge boson, two transverse scalars and one two-form. Using the fact that the transverse scalar in the abelian theory is invariant under the S-duality, we have found that the amplitude satisfies the S-dual Ward identity.
 
\section{Six gauge bosons amplitude}

In this section we are going to show that the scattering amplitudes of only gauge bosons also satisfy the S-dual Ward identity. The first non-zero scattering amplitude of gauge bosons in the DBICS theory is the S-matrix element of four gauge bosons which has only contact terms. It has been shown in \cite{Garousi:2011vs} that this amplitude can be written in manifestly S-dual invariant form as
 \beqa
\cA&\sim& \Tr(\cF_1^T\cM_0\cF_2\cF_3^T\cM_0\cF_4)+\Tr(\cF_1^T\cM_0\cF_3\cF_2^T\cM_0\cF_4)+\Tr(\cF_1^T\cM_0\cF_4\cF_2^T\cM_0\cF_3) \nonumber
\eeqa
All higher point functions have both contact terms as well as massless pole.

The S-matrix element of six gauge bosons in which we are interested in this section, is given by the following Feynman rule:
\beqa
\cA &=&\sum_{i=1}^{10}\cA_i +V(F_1,F_2,F_3, F_4,F_5,F_6)\labell{amp3}
\eeqa
where the massless poles are given as 
  \beqa
{\cal A}_{{1}} &=&V^a(F_1,F_2,F_3, A)G_{ab}(A)V^b(A,F_4,F_5,F_6)\\
{\cal A}_{{2}} &=&V^a(F_1,F_2,F_4, A)G_{ab}(A)V^b(A,F_3,F_5,F_6)\nonumber\\
{\cal A}_{{3}} &=&V^a(F_1,F_2,F_5, A)G_{ab}(A)V^b(A,F_3,F_4,F_6)\nonumber\\
{\cal A}_{{4}} &=&V^a(F_1,F_2,F_6, A)G_{ab}(A)V^b(A,F_3,F_4,F_5)\nonumber\\
{\cal A}_{{5}} &=&V^a(F_1,F_3,F_4, A)G_{ab}(A)V^b(A,F_2,F_5,F_6)\nonumber\\
{\cal A}_{{6}} &=&V^a(F_1,F_3,F_5, A)G_{ab}(A)V^b(A,F_2,F_4,F_6)\nonumber\\
{\cal A}_{{7}} &=&V^a(F_1,F_3,F_6, A)G_{ab}(A)V^b(A,F_2,F_4,F_5)\nonumber\\
{\cal A}_{{8}} &=&V^a(F_1,F_4,F_5, A)G_{ab}(A)V^b(A,F_2,F_3,F_6)\nonumber\\
{\cal A}_{{9}} &=&V^a(F_1,F_4,F_6, A)G_{ab}(A)V^b(A,F_2,F_3,F_5)\nonumber\\
{\cal A}_{{10}} &=&V^a(F_1,F_5,F_6, A)G_{ab}(A)V^b(A,F_2,F_3,F_4)\labell{poF}\nonumber
\eeqa
where the vertex and propagator appear in \reef{poro}, and the contact term can be read from \reef{DBICS} to be
\beqa
V(F_1,F_2,F_3, F_4,F_5,F_6)&\!\!\!\sim\!\!\!&\frac{e^{-3\phi_0}}{12}\bigg[\Tr(F_1\inn F_2\inn F_3\inn F_4\inn F_5\inn F_6)-\frac{3}{8}\Tr(F_1\inn F_2)\Tr(F_3\inn F_4\inn F_5\inn F_6)\nonumber\\
&& +\frac{1}{32}\Tr(F_1\inn F_2)\Tr(F_3\inn F_4)\Tr(F_5\inn F_6)\bigg]+P(1,2,3,4,5,6)\labell{amp4}
\eeqa
where $P(1,2,3,4,5,6)$ stands for all other permutations of $1,2,3,4,5,6$. Note that the action \reef{DBICS} has the coupling $C_0\wedge F\wedge F$, however, for constant background field $C_0$ it does not produce the vertex $V(C_0,F_1,A)^a$, so there is no other term in the amplitude \reef{amp3}.
 
Using the vertex and the propagator in \reef{poro}, one can easily calculate the massless poles in \reef{poF}, \eg $\cA_1$ becomes
\beqa
{\cal A}_{{1}}&=&-\frac{e^{-3\phi_0}}{k \inn k }\bigg[ k \inn F_1\inn F_2\inn F_3\inn F_4\inn F_5\inn F_6\inn k -\frac{1}{4}\Tr(F_1\inn F_2)k \inn F_3\inn F_4\inn F_5\inn F_6\inn k \labell{apFFF}\\
&&\,\,\,\,\,-\frac{1}{4}k \inn F_1\inn F_2\inn F_3\inn F_4\inn k \Tr(F_5\inn F_6)+\frac{1}{16} k \inn F_3\inn F_4\inn k \Tr(F_1\inn F_2) \Tr(F_5\inn F_6)\bigg]+\cdots\nonumber
 \eeqa
where $k=k_1+k_2+k_3=-(k_4+k_5+k_6)$, and dots represent all other terms resulting from the permutations of the labels in the vertex \reef{poro}.  Similar relations can be found  for all other $\cA_2,\cdots , \cA_{10}$. Since the result is in terms of field strength of the external polarization vectors, \ie $F_i$, the amplitude \reef{amp3} satisfies the Ward identity corresponding to the abelian gauge transformation. In the next subsection, we show that it also satisfies the Ward identity corresponding to the S-duality.

\subsection{S-dual Ward identity}

The contact terms \reef{amp4} and the massless poles \reef{apFFF} show that the amplitude \reef{amp3} is independent of the background R-R scalar field $C_0$.   Therefore, to study the S-dual Ward identity of the amplitude \reef{amp3}, one can set $C_0=0$ in the S-duality transformations \reef{BF} and \reef{MM}. To simplify this study, we choose the $SL(2,R)$ matrix $\Lambda$ to be 
\beqa
\Lambda= \pmatrix{0&1 \cr 
-1&0} 
\eeqa
This simplifies the S-duality transformations \reef{BF} and \reef{MM} to
\beqa
F\rightarrow e^{-\phi_0}*F&;&e^{-\phi_0}\rightarrow e^{\phi_0}\labell{sim}
\eeqa
 Using the identity \reef{zero2n}, one finds the following transformations: 
\beqa
 e^{-3\phi_0}\Tr(F_1\inn F_2\inn F_3\inn F_4\inn F_5\inn F_6)&\rightarrow &-e^{-3\phi_0}\Tr(F_1\inn F_2\inn F_3\inn F_4\inn F_5\inn F_6) \nonumber\\
e^{-3\phi_0}\Tr(F_1\inn F_2)\Tr(F_3\inn F_4\inn F_5\inn F_6)&\rightarrow & -e^{-3\phi_0}\Tr(F_1\inn F_2)\Tr(F_3\inn F_4\inn F_5\inn F_6)\nonumber\\
e^{-3\phi_0}\Tr(F_1\inn F_2)\Tr(F_3\inn F_4)\Tr(F_5\inn F_6)&\rightarrow & -e^{-3\phi_0}\Tr(F_1\inn F_2)\Tr(F_3\inn F_4)\Tr(F_5\inn F_6)
\eeqa
As a result, the contact term \reef{amp4} is antisymmetric under the  S-duality transformations \reef{sim}, \ie 
\beqa
V(F_1,F_2,F_3, F_4,F_5,F_6)&\rightarrow &-V(F_1,F_2,F_3, F_4,F_5,F_6)
\eeqa
This is consistent with the fact that the DBI action is not invariant under the S-duality. Because of the above transformation, one expects the massless poles transform   to the massless poles and twice the contact term $V(F_1,F_2,F_3, F_4,F_5,F_6)$. 

The scattering amplitude \reef{amp3} transforms under the S-duality  \reef{sim} to
\beqa
\tilde{\cA} &=&\sum_{i=1}^{10}*\cA_i -V(F_1,F_2,F_3, F_4,F_5,F_6)\labell{amp5}
\eeqa
where $*\cA_1$ is the same as \reef{apFFF} in which $F_i$ is replaced by $*F_i$, and similarly for all other $*\cA_2,\cdots , *\cA_{10}$. 
One may write the $*F_i$ in the amplitude \reef{amp5} as $(*F_i)_{ab}=\frac{1}{2}\eps_{abcd}F_i^{cd}$ and then use the identity \reef{zero2n} to remove the six volume forms, \ie rewrite the amplitude in term of various contractions of $F_i$, and then compare the result with the original amplitude \reef{amp3}. In this approach, in order to use the identity \reef{zero2n}, there are different ways to pair two volume forms. Different pairings give different expressions for the amplitude \reef{amp5} in terms of $F_i$. However, they all must be identical. In fact,   the expression $N\inn *F_1\inn *F_2\inn *F_3\inn *F_4\inn *F_5\inn *F_6\inn M$ where $N,M$ are two arbitrary vectors, can be written in 15 different expressions. These identities result from 15 different paring of volume forms. Using these identities, we have found that the amplitudes \reef{amp3} and \reef{amp5} are identical.  

Alternatively, one may use the matrix form of field strengths $F$ and $*F$
\beqa
F =\pmatrix{0 & E_x & E_y & E_z \cr
-E_x & 0 & -B_z & B_y \cr
-E_y & B_z & 0 & -B_x \cr
-E_z & -B_y & B_x & 0}
\,\,,\,\,
*F =\pmatrix{0 & B_x & B_y & B_z \cr
-B_x & 0 & E_z & -E_y \cr
-B_y & -E_z & 0 & E_x \cr
-B_z & E_y & -E_x & 0}
\eeqa
to write the amplitudes \reef{amp3} and \reef{amp5} in terms of electric field   $\stackrel{\rightarrow}{E}$ and magnetic field $\stackrel{\rightarrow}{B}$. In this way also we   have found that the amplitude \reef{amp3} is invariant under the S-duality \reef{sim}, \ie
\beqa
\tilde{\cA} &=&\cA
\eeqa
This ends our illustration of the fact that the S-matrix element of six gauge bosons satisfies the Ward identity corresponding to the S-duality. We have also calculated the S-matrix element of two gauge bosons and four transverse scalars and found that the amplitude satisfies the Ward identity corresponding to the S-duality transformations \reef{sim}.

 {\bf Acknowledgments}: H.B-A would like to thank D. Mahdavian Yekta  for useful  discussions.


\end{document}